\documentclass{ws-procs9x6}

\def\be{\begin{equation}}
\def\ee{\end{equation}}

\begin{document}

\title{Alternatives to the seesaw mechanism
}

\author{A.~Yu. SMIRNOV
\footnote{\uppercase{T}alk given at the \uppercase{C}onference 
``\uppercase{S}eesaw \uppercase{M}echanism and \uppercase{N}eutrino 
\uppercase{M}asses: 25 \uppercase{Y}ears \uppercase{L}ater'', 10-11 June 2004, 
\uppercase{P}aris.} 
}

\address{International Centre for Theoretical Physics, \\
Strada Costiera 11, \\ 
34014 Trieste, Italy\\ 
E-mail: smirnov@ictp.trieste.it}

\maketitle

\abstracts{The observed pattern of lepton mixing 
does not give an evidence of the seesaw. 
It is easier to disprove seesaw showing that one of the alternatives 
gives dominant contribution to the neutrino mass.  
We consider alternative mechanisms based on (i) small 
(tree level) effective couplings, (ii) small VEV, 
(iii) radiative  generation of masses, (iv)
protection by SUSY breaking scale or by $\mu$-term,  (v) 
small overlap of wave functions of the left and right handed 
neutrino components  in extra dimensions. Seesaw can be the  mechanism 
of suppression of the Dirac mass terms and not dominant mechanism of the 
neutrino mass giving just a sub-leading contribution. 
}

\section{What is wrong with the  Seesaw?}

Needless to say, the seesaw~\cite{sees1,sees,sees2} 
is the most appealing mechanism 
of small neutrino mass generation.
We admire its
simplicity, elegance and naturalness.
We (at least some of us) admire, and doubt, and the reasons for doubt can 
be summarized in the 
following way. \\

\noindent
1). {\it No clear evidence of the seesaw} is seen in the observed pattern 
of  neutrino masses and lepton mixing.

Such an evidence would be obtained if, {\it e.g.}, the 
``HDM + solar SMA MSW '' scenario advocated in 
90ties 
is realized. {\it A priory} this scenario implied
nearly quadratic neutrino mass hierarchy: $m_{\nu} \sim m_u^2$ 
and small (similar to quark) mixing.
The heaviest neutrino was in the eV-range providing the hot component
of the dark matter (HDM) in the Universe with
substantial contribution to the energy density balance.
If this is realized we would agree that the seesaw works, 
the right handed neutrino masses are at the intermediate mass scale
($10^{10} - 10^{12}$ GeV) and there are similar structures of the Dirac
matrices of neutrinos, charged leptons  and quarks.

Instead, the bi-large lepton mixing 
and weak (or none) neutrino mass hierarchy have been found.
Generically, the seesaw  does not reproduce the observed
pattern. For this one needs
(i) some tuning of parameters;
(ii) particular structures of the RH neutrino mass matrix
(very strong hierarchy, off-diagonal elements dominance, {\it etc.}.),.   
or the Dirac mass matrix which differs from the matrices of the charged 
leptons and quarks. 

Though the data do not exclude the seesaw: both
the large mixing and weak mass hierarchy can be reproduced  by 
seesaw~\cite{senhan}.\\

\noindent
2). {\it Even more doubts} in the seesaw will appear if
(i) light sterile neutrinos are found, {\it e.g.}, if MiniBOONE confirms
the LSND result; or  (ii) neutrino mass spectrum turns out to be 
quasi-degenerate. 
Again, both these features can be accommodated in the seesaw.

In fact, even without introduction of new fields one can
supply the seesaw with an additional symmetry which leads
to the three light active neutrinos and one light right handed (sterile) 
neutrino~\cite{rabi}.  

As far as the degenerate spectrum is concerned, one can use the
seesaw type-II with,  {\it e.g.}, SO(3) flavor symmetry\cite{so3}.
Also various symmetries can lead to the degenerate spectrum in the
context of the seesaw type-I.
The double seesaw~\cite{casc}  may reproduce the 
degenerate spectrum. Indeed, let us consider 
(in addition to the RH neutrinos)  three SO(10)
singlets $S$ and the mass matrix of the form
\be
m = 
\left(
\begin{array}{ccc}
0 & m_D & 0\\
m_D & 0 & M_D\\
0 & M_D & M
\end{array} 
\right), 
\label{matr}
\ee
where $ m_D \ll  M_D \ll  M$.
It leads to the light neutrino mass matrix
\be
m = m_D M_D^{-1} M  M_D^{-1 T} m_D^T. 
\label{dss}
\ee
Assume that 
\be
M_D = A m_D,    ~~~ M = M_0 I , 
\label{}
\ee
where $I$ is the unit matrix and $A$ is a constant, that is, the heavy 
and light  Dirac mass 
matrices are proportional each other (the Yukawa couplings of $S$  
follow the family structure). Then the mass matrix  becomes
\be
m_{\nu} = M_0 A^{-2} {I}.
\label{mm}
\ee
Small deviation from this structure gives  small split
of masses and mixing.\\

\noindent
3). {\it No way to prove}... 
What are signatures of the seesaw? 
One can mention 
the neutrinoless double beta decay, leptogenesis,   
flavor changing decays.  
Indeed, discovery of the $\beta\beta_{0\nu}$ decay
will be in favor of seesaw.  
However, this is neither necessary (due to possible cancellations), nor  
sufficient: the positive result of 
the $\beta\beta_{0\nu}$ searches 
is not the prove that the seesaw is the main mechanism  of neutrino mass
generation.

Leptogenesis: it is difficult to establish either.

In the SUSY context the seesaw leads to lepton violating decays,
additional contributions to EDM,  {\it etc.}. However, those provide 
indirect  tests which rely on a number of assumptions 
(see general discussion in \cite{dav}). 

True signature of the seesaw is  detection of the RH Majorana neutrinos,   
measurements of their masses and  couplings with the $W$-bosons. 
Particular versions of the low scale 
seesaw can be tested, in principle, in high energy accelerator 
experiments.  

Apparently it is easier to disprove the  seesaw as the dominant
mechanism of the neutrino mass  generation. \\

\noindent
4). {\it Unpredictable neutrinos:}  
As it was already in the past, neutrinos may
not follow our prejudices
about simplicity, elegance and naturalness... \\

\noindent
5). {\it String theory} tells us that 

(i) Majorana neutrinos are not particularly favored; 

(ii) Appearance of the 126-plets in SO(10) is problematic; 

(iii) Dirac masses can be very small:  

- some  selection rules may exist which lead to small masses;

- values of the Yukawa couplings can be in huge range as 
the  ``Landscape paradigm'' admits; 

- singular structures of the Yukawa matrices may appear.\\

\noindent
6). {\it In a more general context.}..
One can put the  seesaw in some  general context  and then 
argue {\it pro and contra} the context itself.

An example: the seesaw with $M_R = (10^{8} - 10^{15})$ GeV in  SUSY 
or SUSY GUT. 
It can lead to leptogenesis at  temperatures $T > 10^{8}$ GeV.
Inflation should occur at higher temperatures, but in this case the gravitino
problem appears. Furthermore, 
SUSY GUT's have problems with proton decay, FCNC, {\it etc.}.  
Another example is  the  consistent anomaly 
mediation~\cite{mu}. 


\section{Why alternatives?}

\noindent
{\bf Leading or sub-leading.} The effective operator~\cite{eff}
\be
\frac{\lambda_{ij}}{M} (l_i H)^T (l_j H), ~~~~ i,j, = e, \mu, \tau , 
\label{eff}
\ee
where $l_i$  and $H$ are the leptons and Higgs doublets correspondingly, 
generates the Majorana neutrino mass 
$m_{ij} = \lambda_{ij} \langle H \rangle^2/M$. 
For $M = M_{Pl}$ and $\lambda_{ij} \sim 1$ 
it gives $m_{ij} \sim 10^{-5}$ eV.
Such a small contribution is still relevant for phenomenology~\cite{planck}.
Sub-dominant structures of the neutrino mass matrix can be generated 
by the Planck scale interactions~\cite{BV}. So,  the neutrino mass matrix 
can obtain substantial 
contributions from new physics at all possible scales from the EW to 
the Planck scale and from various mechanisms.
We can write the following ``superformula'' for neutrino masses:
\be
m_{\nu} = \sum m_{seesaw} + m_{triplet} + \sum m_{radiative} + m_{SUSY} +
m_{Planck} + ..., 
\ee
where in order the  terms correspond to contributions  from 1).  the seesaw 
realized at different energy 
scales, 2).  the Higgs triplets,  
3). one, two,  {\it etc.} loops effects, 4). SUSY contributions,   
5). the Planck scale physics, {\it etc.}.

One can imagine two possibilities:
(i). The seesaw gives leading contribution, whereas other mechanisms 
produce  sub-leading effects. 
(ii).  The seesaw may turn out to be the sub-leading mechanism.\\

\noindent
{\bf Questions to alternatives.}
There are two questions to any alternative to the seesaw:

\begin{itemize}

\item

Where are the RH neutrinos, $\nu_R$?

\item

If $\nu_R$ exist, why the Dirac mass terms effects are small or absent?

\end{itemize}

There are two possible answers to the second question:

1).  The Dirac masses are forbidden by symmetry with  
immediate objection that this is unnatural - why neutrino but   not other 
fermion masses are suppressed?

2). Dirac mass contributions are suppressed by couplings with the heavy 
degrees of freedom. Here again one can consider two 
possibilities:

(i). Introduction of  large Majorana mass of the RH neutrinos. In this 
way we come back to the seesaw.
So, the seesaw can be the mechanism
of suppression of the Dirac mass term and  not the main  contribution to
neutrino mass; 

(ii). Introduction of  another (large)
Dirac mass terms formed by $\nu_R$ and new singlet $N$. 
In the basis $(\nu, \nu_R, N)$  consider the mass matrix 
\be
m =
\left(
\begin{array}{ccc}
0 & m_D & 0\\
m_D & 0 & M_D\\
0 & M_D & 0
\end{array}
\right),
\label{multi}
\ee
which leads to one strictly massless neutrino \cite{multsin}. For
$m_D \ll M$ the admixture of the heavy lepton is negligible.
We will refer to  this possibility as 
to the {\it multi-singlet} mechanism of the suppression.\\ 

\noindent
{\bf General context.} A general context for consideration of neutrino masses 
can be formulated in the following way. 
Beyond the Standard Model 
there are three RH neutrinos,  $\nu_{Rj}$, and also
a number of other SM singlets,  $S_i$.
The Yukawa couplings of these singlets with the active neutrinos, 
\be
h_{kj} \bar{l}_k \nu_{Rj} H + f_{ik} \bar{l}_k S_{i} H, 
\label{ycoup}
\ee
are small due to symmetry, or their contributions to neutrino 
masses are suppressed 
by the seesaw 
or by ``multi-singlet'' mechanism.\\

\noindent
{\bf Prove or Disprove.}
It is easier to test alternatives to the seesaw than the  seesaw itself.
In fact, a number of  alternatives is related to new physics
at the electroweak scale which can be tested at  high energy accelerators.
Therefore the  way to proceed is to exclude alternatives.
Though it may  not be possible to exclude all of them, 
still more confidence in the seesaw  will be obtained. 
Or it may happen that 
validity of the   alternative will be proven. In this case
the seesaw still can play a role of the sub-leading mechanism and
we will put bounds on its contribution.

So,  we need to  consider the alternatives because 
this is probably the most efficient way to disprove or prove the seesaw.

\section{Classifying alternatives}

There  are  different ways to classify the alternatives. One can use

\begin{itemize}

\item
diagrams (tree level, one loop, two loops, {\it etc.}) \cite{cl-d};

\item
possible field operators which lead to the neutrino masses~\cite{cl-op};

\item
physical context.

\end{itemize}
Let us consider  the latter possibility.
The first step in classification  is the nature of the neutrino mass 
terms: Dirac or Majorana and their gauge symmetry properties.
In the Majorana case (weak isotriplet) the mass can be generated by the 
effective operator 
which includes interaction with two Higgs doublets or with Higgs triplet. 


In turn, smallness of the mass can be due to  
(i) small VEV; (ii) small (effective) coupling 
which may appear at tree level or 
radiatively; (iii)  small overlap  of wave functions
of the left and right handed neutrino components  
in extra dimensions. \\

\noindent
{\bf Small Yukawa couplings.} The observed neutrino masses can be 
reproduced if
$
h_{ij} \sim 10^{-13}
$
in (\ref{ycoup}). For usual  Dirac type Yukawa couplings
similar to the quark or charged lepton couplings these values
look very unnatural and require some explanation.

One can consider the following scenario: the 
usual Yukawa couplings for the $\nu_L$ and  $\nu_R$ are not small
(of the same size as quark and lepton couplings).  However the
corresponding masses are strongly suppressed by
the seesaw or multi-singlet mechanisms.

Neutrino  masses which we observe in the oscillation experiments are
formed by $\nu_L$ and new singlets, $S$,  
(see second term in (\ref{ycoup})) 
which have no analogy in the quark sector.  
These singlets may have some particular symmetry properties or/and
come from the hidden sector of theory. 
As a consequence, their couplings, $f_{ij}$,  
can be small.

Clearly, scenario with small Dirac couplings will be excluded if the
neutrinoless double beta decay is discovered and it will be 
shown that the decay is due to light Majorana neutrinos.\\

\noindent
{\bf Small effective couplings.} Non-renormalizable operators
\be
a_{ij} \bar{l}_{i} S_j H \frac{S}{M}
\ee
can generate small effective Yukawa couplings  
\be
h_{ij} = a_{ij} \frac{\langle S \rangle}{M}   
\ee
for  $a_{ij}  \sim O(1)$,  if  
$\langle S \rangle / M \sim 10^{-13}$.
(Renormalizable coupling can be suppressed by symmetry).
One can consider the SUSY or GUT scales for $M$, if $\langle S \rangle$ 
is
at the electroweak scale, or take $m_{3/2}/ M_{Pl}$. Another possibility 
is to assume a small VEV of $S$ \cite{effcoup}.


Hierarchy $\langle S \rangle / M$   can be substantially reduced if the 
effective coupling
appears in higher order non-renormalizable interactions:
\be
h_{ij} = a_{ij} \frac{ \Pi_{k = 1...n} \langle S_k \rangle}{M^n}.
\ee\\

\noindent
{\bf Higgs triplet mechanism.} 
The  Majorana neutrino mass can be  generated at tree level
by  coupling with the Higgs triplet~\cite{tripl,tripl1,ma,ma1}
$\Delta \equiv (\Delta^{++}, \Delta^{+}, \Delta^{0})$: 
$
g_{\alpha \beta} l_{\alpha}^T l_{\beta} \Delta . 
$

The electroweak precision measurements give $\langle \Delta \rangle /
\langle H \rangle < 0.03$. To avoid appearance of the triplet 
Majoron~\cite{tripl1} the 
coupling $\mu \Delta H H$ with the Higgs doublets should be introduced.
If $g_{\alpha  \beta} \sim 1$, then $\langle \Delta^0 \rangle \sim 1$ eV.

Various scenarios depend on the  triplet mass $M_{\Delta}$.
If $M_{\Delta}, \mu \gg \langle H \rangle$, the induced VEV appears 
$\langle \Delta^0 \rangle \sim \langle H \rangle^2 
\mu/M_{\Delta}^2$~\cite{tripl} 
and we arrive at the seesaw type-II. 
If in contrast, $M_{\Delta} \sim \langle H \rangle$ and $\mu \ll \langle H 
\rangle$, we find $\langle \Delta^0 \rangle \sim \mu$.
The  pseudo-Majoron mass $\sim \mu \langle H \rangle^2 /\langle \Delta^0 
\rangle$ can be made large enough to avoid 
the experimental bounds, in particular, from measured $Z^0$ 
width~\cite{ma,ma1}. 

One can consider the effective coupling of  
neutrinos with triplet which arises  from   the non-renormalizable 
interactions:
\be
g_{\alpha \beta}  \frac{S}{M} l_{\alpha}^T l_{\beta} \Delta , 
\label{tri3}
\ee
where the singlet $S$ acquires  VEV. This allows us to increase 
the required VEV of $\Delta$. Another possibility apears in  
models with the triplet and two Higgs doublets~\cite{ma1}.  

\section{Mechanisms never die}

\begin{figure}[ht]
\centerline{\epsfxsize=3.5in\epsfbox{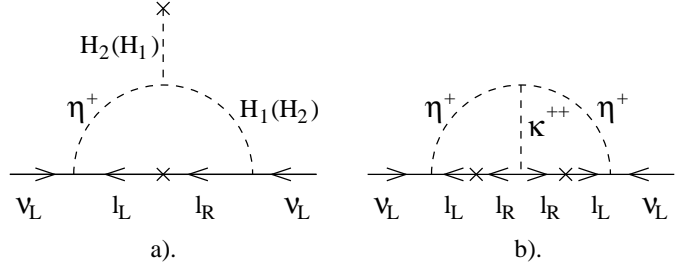}}   
\caption{Diagrams of the radiative  neutrino mass generation.
a). Zee mechanism. b). Babu-Zee mechanism. Flavor indices
are omitted.
\label{rad}}
\end{figure}

\noindent
{\bf Zee mechanism.} 
There is no RH neutrinos,  instead new scalar bosons are introduced:
the charged singlet of SU(2), $\eta^+$,  and second Higgs doublet
$H_2$. Their couplings 
\be
l^T \hat{f} i\sigma_2 l \eta^+ +
\sum_{i = 1,2} \bar{l} \hat{f}_i e H_i , 
\label{coup}
\ee
where $\hat{f}_i$ is the matrix of the Yukawa couplings of Higgs $H_i$ 
(i = 1,2), 
generate neutrino masses in one loop~\cite{zee} (fig. \ref{rad} a) 
\be
m_{\nu} = A [(\hat{f} \hat{m}^2 + \hat{m}^2 \hat{f}^T) -
v (\cos \beta)^{-1}(\hat{f} \hat{m} \hat{f}_2 + \hat{f}_2^T \hat{m} 
\hat{f}^T)]. 
\label{massZ}
\ee
Here $A = \sin 2\theta_Z \ln(M_2/M_1)/(8 \pi^2 v \tan\beta))$,
$\hat{m} = diag(m_e, m_{\mu}, m_{\tau})$, $\tan\beta \equiv v_1/v_2$,
$v^2 \equiv v_1^2 + v_2^2$.
In the minimal version only one Higgs doublet couples to
leptons, $\hat{f}_2 = 0$, and consequently, only the first term in (\ref{massZ}) 
contributes to the mass. 
The neutrino mass matrix has zero diagonal elements 
and therefore experimentally excluded~\cite{zeeH}:  
it  can not reconcile two large mixings,
one small mixing and hierarchy of $\Delta m^2$. 

There are several ways to revive the Zee-model~\cite{zeeH}.

1). Introduction of the non-zero couplings of both Higgs doublets
with leptons (non-zero second term in (\ref{massZ})) leads to 
non-zero diagonal mass terms of  the mass matrix and a possibility to 
describe all experimental results. The model predicts 
decays $\tau \rightarrow \mu \mu \mu$,
$\mu \rightarrow eee$, $\tau \rightarrow \mu \mu e$ due 
to the Higgs exchange with the branching ratios  being 
 2 - 3 orders of magnitude below the present 
experimental bounds.

2). Some other mechanism can give additional contributions 
to the neutrino mass matrix, in particular, to the 
diagonal terms: Higgs triplet, scalar singlet, two loop 
contribution~\cite{zeeH}.

3). Additional contributions  to the mass matrix appear 
if new leptons, in particular, sterile neutrinos,  exist.

The model is testable in the precision electroweak measurements,
searches for charged Higgses and  rare decays.

Still the problem exists with explanation of smallness
of the  couplings: the neutrino data require
inverse hierarchy of $f_{\alpha\beta}$ and 
$f_{\alpha\beta} \sim 10^{-4}$.\\

\noindent
{\bf Zee-Babu mechanism.}
There is no RH neutrinos. New scalar bosons, singlets of SU(2), 
$\eta^+$ and $k^{++}$, are introduced with the following couplings
\be
l^T \hat{f} l \eta^+  + l_R^T \hat{h} l_R k^{++}. 
\ee
Here $\hat{f}$ and $\hat{h}$ are the matrices of Yukawa couplings in the 
flavor basis.
The Majorana neutrino masses are generated in two loops~\cite{zee,babu} (fig. 
\ref{rad}b):
\be
m_{\nu} \sim 8 \mu \hat{f} \hat{m}_l \hat{h} \hat{m}_l \hat{f} I, 
\ee
where $\hat{m}_l \equiv diag(m_e, m_{\mu}, m_{\tau})$.

The main features of the model (see~\cite{babuR,yasue} 
for recent discussion) are: 
one  massless neutrino; inverted hierarchy of the 
couplings in the flavor basis; values of the couplings:
$f, h \sim 0.1$.
The model is  testable: new charged scalar bosons
exist at the electroweak scale, the decay rates for
$\mu \rightarrow e \gamma$, and $\tau \rightarrow 3\mu$ are within 
a reach of the forthcoming experiments.\\

\noindent
{\bf R-parity violating SUSY.} Terms of the superpotential
\be
W = - \mu_{\alpha} l_{\alpha} H_u - 0.5 \lambda_{\alpha \beta m}
+ \lambda'_{\alpha n m} l_{\alpha} Q_n d^c_m  + h_{m n} H_u Q_n u^c_m
\label{sup}
\ee
violate  the lepton number. 
Here $\alpha  = 0, 1, 2, 3$, $l_0 \equiv H_d$, $m = 1, 2, 3$. No RH neutrinos
are introduced.

\begin{figure}[ht]
\centerline{\epsfxsize=3.1in\epsfbox{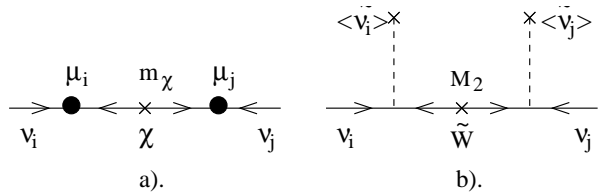}}   
\caption{Diagrams for the neutrino mass generation at tree level in the 
model with the R-parity violation. a) In the basis 
where $\langle \tilde{\nu} \rangle 
= 0$.  b). In the basis where $\mu_m = 0 (m = 1,2,3)$.  
\label{rpvt}}
\end{figure}

The bi-linear terms in (\ref{sup}) \cite{bili}   give the dominant
tree level contribution to neutrino  masses. In the basis where sneutrinos have 
zero VEV's, 
$\langle \tilde{\nu} \rangle = 0$,  the  masses are produced
via mixing with Higgsinos 
by the diagram (fig.~\ref{rpvt}a):
\be
m_{ij} = \mu_i \mu_j  \frac{\cos^2 \beta}{m_{\chi}}. 
\ee
In the basis with $\mu_m = 0$, the neutrino masses are
generated by the electroweak seesaw:  light neutrinos are mixed with
wino (neutralino) after sneutrinos get VEV's (fig.~\ref{rpvt}b): $m_{ij} 
=
A \langle \tilde{\nu}_i \rangle \langle \tilde{\nu}_j \rangle$.
Here $A = h_b^2 /(16 \pi^2 m^2_{\tilde W})$.

Only one neutrino acquires mass at this tree level (in assumption of
universality of the soft symmetry breaking 
terms)~\cite{bili,Josh,review,grossman,diaz}. Moreover,  mixing is determined by 
the ratios  of the mass parameters: $\mu_i/\mu_j$.

\begin{figure}[ht]
\centerline{\epsfxsize=3.1in\epsfbox{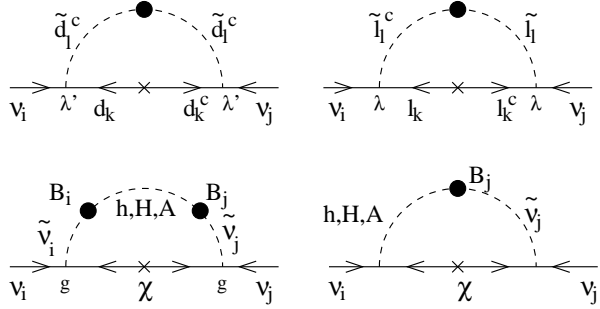}}   
\caption{One loop diagrams for the neutrino mass generation in the model 
with R-parity violation.
\label{rpvl}}
\end{figure}

The trilinear RpV couplings in (\ref{sup}) and soft symmetry
breaking terms  (characterized by the mass parameters
$B_i$) generate one loop contributions 
(fig.~\ref{rpvl}). 
As a result, natural neutrino mass hierarchy appears:
at tree level one (largest) mass as well as one large mixing 
are generated; loop contributions produce other small  masses
and small mixings.

Correct scale of the neutrino masses requires
$\mu_m \sim 10^{-4}$ GeV, which in turn,  implies further
structuring - explanation of the hierarchy
$\mu_m \ll m_0$.

In the original version (the universal SUSY breaking at high scale), 
generically only one large mixing can be obtained. 
Explanation of neutrino data with two large mixings requires
violation of universality of the soft symmetry breaking terms.
Both the Higgs-lepton  $(\mu - \mu_i)$ and flavor universality should be 
broken\cite{grossman,diaz}, that is, $B_{i} \neq B_{j}$  
at the high scale.

The RpV models have very rich phenomenology: new physics at
colliders, relatively fast flavor violating decays,
new neutrino interactions {\it etc.}.

\section{Old and New}

\noindent
{\bf SUSY violation and neutrino 
mass}~\cite{ben-sm,arkani,borzu,abel,arn,russel,casas}.  
It was observed long time ago~\cite{ben-sm} that
\be
m = \frac{m_{3/2} v_{EW}}{M_{Pl}} \sim 10^{-4}~{\rm eV}, 
\label{mass}
\ee
where $m_{3/2} \sim 1$ TeV is the gravitino mass and $v_{EW} \equiv 
\langle H \rangle$ is the
electroweak VEV. 
The value (\ref{mass}) is close to the
scale of observed  neutrino masses and certainly is interesting
from the phenomenology point of view.
It can be generated by the Yukawa interaction
\be
\lambda \bar{l} S H, ~~~~~ \lambda = \frac{m_{3/2}}{M_{Pl}}.
\label{inter}
\ee
It can mix active neutrinos with singlets of SM, {\it e.g.},
form usual mass term, or mix neutrinos with
modulino~\cite{ben-sm}.

The  interaction (\ref{inter}) may follow from  
non-renormalizable term in the superpotential or from the K\"ahler potential 
similarly to appearance of the $\mu$ - term in the  Giudice-Masiero mechanism:
\be
K = \frac{1}{M_{Pl}} P(S, z, z^*) \bar{l} H + h.c., 
\ee
where $z$ are the Wilson lines~\cite{ben-sm,borzu}.

The mass  (\ref{mass}) is too small to explain 
observations, 
but there are various ways to  enhance it.
In general, the mass can be written as
\be
m = \frac{\alpha \eta m_{3/2} v_{EW}}{M_{Pl}}, 
\label{mass1}
\ee
where $\eta$ describes the renormalization group effect and
$\alpha$ is an additional numerical factor.

1). The gravitino mass can be larger: the value $m_{3/2} \sim 10^2$ TeV
brings the mass to the correct range $10^{-2}$ eV.
Such a scale for $m_{3/2}$ appears, {\it e.g.}, in the model of 
``consistent anomaly mediation'' \cite{mu}.

2). One can take $M_{GUT}$ instead of $M_{Pl}$ which leads to
$\alpha \sim 10^2$.

3). Large factor  $\alpha$
may appear as a  consequence of particular mechanism of mass 
generation. 
For instance, the terms in the  K\"ahler potential
\be
K = \frac{1}{M} P(S, \sigma, \sigma^*) \bar{l} H N + h.c.
\ee
may have the cut-off parameter $M = 10^{17}$ GeV -  below the 
Planck mass. Here $\sigma$ are the fields of the hidden sector.
Then the  dominant contribution to the neutrino mass is given by~\cite{abel}
$
v_{EW} F_{\sigma}/M^2, 
$
where $F_{\sigma} = \sqrt{3} M_{Pl} m_{3/2}$. 
It leads to a correct range 
\be
m = \frac{\eta m_{3/2} v_{EW}}{M} \frac{\sqrt{3} M_{Pl}}{M} \sim 0.05 ~~{\rm
eV}. 
\label{mass2}
\ee\\

\noindent
{\bf Variation on the theme.} 
Small Dirac type Yukawa couplings can appear from the superpotential,
whereas the Majorana masses follow from the K\"ahler potential \cite{russel}:
\be
W = g \frac{X}{M} l N H, ~~~~ K = h \frac{Y^*}{M} N N. 
\label{}
\ee
Here $M = M_{Pl}/\sqrt{8\pi}$,  $X$ and $Y$ are the fields of hidden sector, 
with  
VEV's
$\langle X_A \rangle = m_I$ and $\langle Y_F \rangle = m_I^2$ at the
intermediate mass 
scale $m_I = \sqrt{m_{3/2} M_{Pl}}$.
Then 
$
m_D = g v_{EW} \sqrt{m_{3/2}/M}$, $m_N = h m_{3/2}$ 
are small, and the TeV-scale seesaw gives 
$m_{\nu} \sim v_{EW}^2/M g^2/h \sim 10^{-3}$ eV.

\begin{figure}[ht]
\centerline{\epsfxsize=1.5in\epsfbox{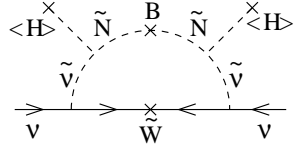}}   
\caption{Diagram for the neutrino mass generation in the model$^{36}$.
\label{newd}}
\end{figure}

If in addition the term in the K\"ahler potential
\be
K = ...+ h \frac{Y^* Y X^*}{ M^3} N N
\ee
is introduced~\cite{russel}, the main contribution, $\sim 0.05$ eV, follows from 
the one loop diagram shown in fig.~\ref{newd}.\\

\noindent
{\bf $\mu$-term mixing.} 
Small Dirac mass of neutrino can be related to (protected by) a 
small value  of the $\mu$ - parameter in the context of SUSY models 
and not to the EW scale~\cite{kitano} or $m_{3/2}$. 
Such a possibility is realized in the SU(5) GUT with R-symmetry~\cite{kitano}.
The following R-charges for the matter and Higgs fields  are prescribed 
(second number in the bracket):
\be
\bar{F} (\bar{5}, 1), ~H(10, 1),~ N(1, - 1),~~
H(5,0), ~\bar{H}(5, 0), ~ H'(5, 2), ~\bar{H'}(5,0).  
\ee
Notice that the R-charge of the RH neutrinos differs from the 
charges of  other 
matter fields.  Also new Higgs multiplets  ($H', \bar{H'}$) are introduced
and one of them has non-zero R-charge.
The superpotential includes  
\be
W = f \bar{F} H' N + M_H H' \bar{H}' + ... . 
\label{wterm1}
\ee
The $\mu$-term as well as the Majorana mass terms for the RH
neutrinos are forbidden by the R-symmetry.

SUSY breaking leads to the $R$-symmetry breaking  
and can  generate the following operators at the TeV scale:
\be
W_R = \mu H \bar{H} + \mu' H \bar{H}'. 
\label{wterm2}
\ee
The last terms in (\ref{wterm1}) and (\ref{wterm2})
mix  $H$ and $H'$ with the angle 
$\sim \mu'/ M_H$.  Consequently, the first term in (\ref{wterm1}) 
generates  small Yukawa coupling: $(f\mu'/M_H) \bar{F}H N$.
As a result,  neutrinos acquire the Dirac masses   $f \mu' 
v_{EW}/M_H$.

\section{Extra dimensions and Extra possibilities}

Theories with extra space dimensions provide qualitatively new mechanism
of generation of the small {\it Dirac} neutrino mass. There are 
different
scenarios, however their common feature 
can be called the overlap suppression: the 
overlap of wave functions of the left, 
$\nu_L(y)$, and right , $\nu_R(y)$ handed  components in
extra dimensions (coordinate $y$). The suppression occurs due to different
localizations 
of the $\nu_L(y)$ and $\nu_R(y)$.
The effective Yukawa coupling is proportional to the overlap. 
One can introduce  also suppression of  overlap of the neutrino and  
Higgs fields. Let us  consider  
realizations of the overlap mechanism  in 
different extra dimensional scenarios.\\

\begin{figure}[ht]
\centerline{\epsfxsize=2.9in\epsfbox{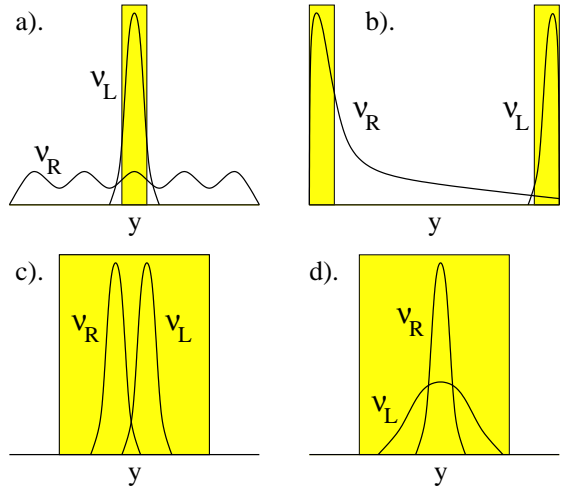}}   
\caption{The overlap mechanism of small Dirac neutrino mass generation in
models with extra spatial dimensions. a). Large flat extra dimensions.
b). Warped extra dimensions. c - d). Models of ``fat'' branes.
\label{extra}}
\end{figure}

\noindent
{\bf ... in large flat extra dimensions.} 
The setup is the $3D$ spatial brane in $(3+\delta)D$ bulk~\cite{add}.
Extra dimensions have large radii $R_i \gg 1/M_{Pl}$ which allows 
one to reduce the fundamental scale of theory
down to $M^* \sim 10 - 100$ TeV~\cite{add}.
The left handed neutrino is localized on the brane, 
whereas the right handed component (being a singlet of the gauge group)
propagates in the bulk (see fig.~\ref{extra}a).

Let us consider one extra dimension of
radius $R$. Since the RH component is not localized,
we find from the normalization condition
that its wave function has typical value  $\nu_{R}(y) \sim 1/\sqrt{R}$. 
The width of the brane is of the order $d \sim 1/M^*$, so the amplitude 
of probability to find the RH neutrino on the brane equals
\be
d^{1/2} \nu_R \sim \frac{1}{\sqrt{M^* R}}. 
\ee
Since the LH neutrino is localized on the brane, this amplitude 
describes  the overlap of the wave functions.
For $\delta$   extra dimensions we get for the overlap
factor
$
1/\sqrt{M^{* \delta} V_\delta}, 
$
where $V_{\delta}$ is the  volume of extra dimensions. 
If $\lambda$ is the Yukawa coupling for  neutrinos in the 
$(4 + \delta)D$ theory, the effective coupling in 4D will be suppressed
by this overlap factor. Consequently, 
\be
m_D = \lambda v_{EW} \frac{1}{\sqrt{M^{* \delta} V_\delta}} = \lambda v_{EW}  
\frac{M^*}{M_{Pl}}, 
\ee
where in the second equality the relation 
$M_{Pl}^2 = M^{*2 + \delta}  V_{\delta}$ have been used. 
For $M^* \sim 100$ TeV and  $\lambda \sim 1$ we obtain 
$m_D \sim 10^{-2}$ eV. \\

\noindent
{\bf ...in warped extra dimensions.}
The setting is one extra dimension compactified on the   
$S^1/Z_2$ orbifold,  and non-factorizable metric.
The coordinate in the extra dimension 
is parameterized by $r_c \phi$, where $r_c$ is the radius of extra 
dimension and the angle $\phi$  changes from 0 
to $\pi$. Two branes are localized
in different points of extra dimension: the ``hidden'' brane at
$\phi = 0$ and the observable one  at $\phi = \pi$~ \cite{RS}. 
The wave function of the  RH neutrino $\nu_R(\phi)$ 
is centered on the hidden brane,  whereas  the LH one -  on the visible
brane (see fig.~\ref{extra} b).
Due to warp geometry  $\nu_R(\phi)$ 
 exponentially decreases from the  hidden to the
observable brane. On the observable brane it is given by
\be
\nu^R(\pi) \sim \epsilon^{\nu -1/2},  ~~~
\epsilon = e^{-k r_c \pi} = \frac{v_{EW}}{M_{Pl}} . 
\label{wfR}
\ee
Here $M_{Pl}$ is the Planck scale, 
$k \sim M_{Pl}$ is the curvature parameter. In (\ref{wfR}) $\nu \equiv  
m/k$ and
$m \sim M_{Pl}$ is the Dirac mass in 5D.
Essentially $\nu_R(\pi)$ gives the overlap factor and the
Dirac mass on the visible brane equals
\be
m_D = \lambda \nu^R(\pi) v_{EW} \sim M \left(\frac{V_{EW}}{M}\right)^{\nu
+ 1/2}. 
\ee
For $\nu = 1.1 - 1.6$ we obtain the mass in the required 
range. 

Different realization when $\nu_R$ is on the TeV brane,  
whereas $\nu_L$ is on the Planck brane, has 
been suggested  in~\cite{ghergetta}. \\

\noindent
{\bf...on the fat brane.}
The  LH and RH neutrino wave functions can be localized differently
on the same ``fat'' brane \cite{fat}. There are various possibilities to
suppress the overlap:

1). localize $\nu_L$ and $\nu_R$ in different places of the brane
(fig. \ref{extra} c);

2). arrange parameters in such a way that  {\it e.g.}  the RH neutrino is 
localized in the
narrow region of the fat brane whereas the LH neutrino wave function is
distributed in whole the brane (fig.\ref{extra}d) \cite{figd}.

\section{Conclusion}

1. Comprehensive tests of the alternatives to the seesaw 
is probably the only way to uncover mechanism of neutrino mass generation. 
A number of alternatives  are related to interesting ideas and 
concepts of physics beyond the SM and therefore  
their tests   provide (often unique) tests of this new 
physics. 

The alternatives  include the radiative mass generation, 
small effective coupling,  small VEV,  mechanisms related to 
SUSY breaking or generation of $\mu$ term, 
the overlap mechanism in extra dimensions, {\it etc.}.

2. It is easy to test alternatives (at least some of them related to 
the EW scale physics) than the seesaw itself. 
And it may happen that one of the  alternatives gives the  
main contribution to the neutrino mass, thus excluding the seesaw
as the dominant mechanism.

3. It is possible that some alternatives  produce  relevant
sub-leading  contributions to the neutrino mass matrix being responsible for
physics ``beyond the seesaw''.



%
%
%
%

\end{document}